\documentclass[a4paper]{jpconf}
\usepackage{iopams}  
\usepackage{graphicx}

\usepackage{amssymb}

  \expandafter\let\csname equation*\endcsname\relax
  \expandafter\let\csname endequation*\endcsname\relax

\usepackage{amsmath,mathtools}

\usepackage{url}
\usepackage{natbib}

\usepackage{appendix}
\usepackage[margin=1cm]{caption}
\usepackage{fullpage}

\usepackage{hyperref}

\usepackage{color}

\usepackage{dsfont}

\newcommand{\Msun}{\ensuremath{M_{ \odot }}}

\begin{document}
\title{The sensitivity of pulsar timing arrays}
\author{C J Moore}
\address{Institute of Astronomy, Madingley Road, Cambridge, CB3 0HA, United Kingdom}
\ead{cjm96@ast.cam.ac.uk}

\begin{abstract}
Simple analytic expressions are derived for the sensitivity curve of a pulsar timing array (PTA) to both a monochromatic source of gravitational waves and an isotropic stochastic background of gravitational waves. These derivations are performed in both frequentist and Bayesian frameworks and the results verified with numerical injections and recovery in mock PTA datasets. The consistency of the frequentist and Bayesian approaches is demonstrated and the results are used to emphasise the fact that the sensitivity curve of a PTA depends not only on the properties of the PTA itself, but also on the properties of the gravitational wave source being observed. 
\end{abstract}

\section{Introduction}
Pulsar timing arrays (PTAs) are ongoing attempts to detect low frequency gravitational waves (GWs) by using the timing of a network of galactic millisecond pulsars. The chosen millisecond pulsars are extremely stable rotators which allow a timing model of sufficient accuracy to be constructed to account for every single rotation of the pulsar over the many years of observation. After subtracting the timing model from the pulse arrival times the remaining timing residuals include any unmodeled effects; e.g. timing noise and (possibly) GWs. All of the pulsars reside in the same GW field and hence the induced timing residuals in different pulsars are correlated by an amount that depends on their angular separation. It is this correlation which allows a network of pulsars spread across the sky to be exploited as a GW detector.

PTAs are sensitive to GWs in the range $10^{-9}\,\textrm{Hz}\lesssim f \lesssim 10^{-6}\,\textrm{Hz}$ and are typically most sensitive to the lowest frequencies in this range. Potential sources of GWs in this frequency band are individual supermassive black hole binaries with chirp masses around ${\cal{M}}\approx10^{9}\Msun$ out to redshifts of $z\approx 2$ (\citep{jaffe-backer-2003,wyithe-loeb-2003}). If there are many sources per unit frequency then they cease to be individually resolvable and instead overlap to form a stochastic background of GWs (\citep{sesana-vecchio-colacino-2008,sesana-vecchio-volonteri-2009,sesana-vecchio-2010}). It is currently unclear whether the first detection of GWs by PTAs will be an individual binary or a stochastic background. Stochastic backgrounds may also be caused by other mechanisms such as cosmic string networks \citep{olmez-2010}, or primordial GWs (\citep{grishchuk-1976,grishchuk-2005}).

There are currently three PTAs operating: the European Pulsar Timing Array (EPTA\footnote{\url{http://www.epta.eu.org/}}, \cite{eptareview2013}), the Parkes Pulsar Timing Array (PPTA\footnote{\url{http://www.atnf.csiro.au/research/pulsar/ppta/}}, \cite{parkesreview2013}) and the North American Nanohertz Observatory for Gravitational Waves (NANOGrav\footnote{\url{http://nanograv.org/}}, \cite{nanogravreview2013}). There are plans to merge the seperate PTAs to form the International Pulsar Timing Array (IPTA\footnote{\url{http://www.ipta4gw.org/}}, \cite{iptareview2013}).

These proceedings are based on a talk given at the 10th International LISA Symposium, this talk in turn was based largely on the work done in \cite{MooreTaylorGair2014}. In Sec.\ \ref{sec:PTA} of this paper the properties of the PTA which will be assumed throughout the remainder of the paper are summarised, Sec.\ \ref{sec:mono} describes the sensitivity of this PTA to an individual binary, Sec.\ \ref{sec:stoch} describes the sensitivity to a stochastic background and final concluding remarks are given in Sec.\ \ref{sec:conclusions}.

\section{Our canonical PTA}\label{sec:PTA}
In order to produce definite values for the sensitivity curves it is necessary to assume particular values for the PTA parameters. For all calculations in this paper the PTA was assumed to consist of 36 pulsars randomly placed on the sky all timed with a cadence of $1/(2\, \textrm{weeks})$ for a total of $5\,\textrm{years}$ with a precision of $100\,\textrm{ns}$. This is comparable to the \textsc{Open1} dataset in the recent IPTA data challenge\footnote{\url{http://www.ipta4gw.org/?page_id=89}}. It should be noted that the performance of this canonical PTA is significantly better than any of the current PTAs.

\section{Sensitivity to an individual binary}\label{sec:mono}
Binary black holes, in contrast to the stochastic backgrounds considered in Sec.\ \ref{sec:stoch}, are a GW source which can be accurately modelled over the entire duration of observations; hence matched templates may be constructed for the purpose of detection and parameter estimation. PTAs are sensitive to individual binary black holes with very large chirp masses ($10^{8}\Msun\lesssim{\cal{M}}\lesssim 10^{10}\Msun$) in the early, adiabatic stage of their inspiral. These systems are typically in circular orbits evolving on timescales longer than the total length of our observations and appear in the data as quasi-monochromatic signals. Therefore, for simplicity, in this section we consider the sensitivity of our PTA to a monochromatic source of GWs as a proxy for an individual binary.

\subsection{Frequentist detection}\label{subsec:freqmono}
The frequentist procedure for claiming a detection of a signal in the PTA dataset is to define a scalar detection statistic, ${\cal{S}}$, which is a function of the dataset. The signal to noise ratio (SNR) of this statistic, $\varrho$, is then defined as the expectation of this statistic in the presence of a signal divided by the square root of the variance of the statistic in the absence of a signal. A detection is claimed if the value of the detection statistic evaluated on the actual measured data exceeds some predetermined threshold, $\varrho_{\textrm{th}}$ (the value of this threshold may be tuned to fix the false alarm rate).

Here the detection statistic was chosen to be the cross correlation of two data streams with the optimal (or ``matched'') filter function, ${\tilde{\bf{K}}}(f)$; see Eq.\ \ref{eq:statistic}. This statistic has the advantage that it gives an SNR which is quartic in the signal amplitude (see Eq.\ \ref{eq:statistic}, we tentatively venture to call this the ``power SNR'' in contrast to the usual ``amplitude'' SNR) and hence may be directly compared with the stochastic background SNR (see Sec.\ \ref{sec:stoch}). It should be noted that in the case of a single binary it is also possible to use the usual matched filter statistic and associated ``amplitude'' SNR for the purposes of detection, but in the case of the stochastic background one is forced to use the cross correlation statistic because the signal in each pulsar is swamped by the intrinsic pulsar red-noise.
\begin{eqnarray}\label{eq:statistic} &{\cal{S}}=\int\textrm{d}t\,\int\textrm{d}t'\,{\bf{s}}(t)^{T}{\bf{K}}^{\dagger}(t-t'){\bf{s}}(t')=\int\textrm{d}f\;\tilde{{\bf{s}}}(f)^{T}\tilde{{\bf{K}}}^{\dagger}(f)\tilde{{\bf{s}}}(f) \\
&\tilde{{\bf{K}}}(f)=\left. \frac{\tilde{{\bf{h}}}(f)\tilde{{\bf{h}}}(f)^{T}}{S_{n}^{2}}\right|_{\textrm{diag}\rightarrow 0}\; \Rightarrow \;\varrho^{2}=\sum_{y}\sum_{x>y}\frac{8}{T}\int\textrm{d}f\;\frac{\left|\tilde{h}_{x}(f)\right|^{2}\left|\tilde{h}_{y}(f)\right|^{2}}{S_{n}^{2}} \, . \nonumber
\end{eqnarray}

In evaluating the SNR in Eq.\ \ref{eq:statistic} the noise terms in the timing residuals from pulsars $x$ and $y$ have been assumed to be white, Gaussian and uncorrelated between pulsars. If pulsar $x$ is timed with precision $\sigma_{x}$ and with a cadence $1/\delta t_{x}$ then this leads to the familiar expression
\begin{equation}\label{eq:noise} \left<\tilde{n}^{*}_{x}(f)\tilde{n}_{y}(f')\right>=\frac{1}{2}S_{n}(f)\delta_{xy}\delta (f-f') \quad \textrm{where} \quad S_{n}=2\sigma_{x}^{2}\delta t_{x} .\end{equation}

The red curve in Fig.\ \ref{fig:mono_comparison} shows the sensitivity curve obtained by setting $\varrho=\varrho_{\textrm{th}}$ in Eq.\ \ref{eq:statistic} and rearranging to find the amplitude as a function of frequency. This sensitivity curve has the obvious problem that it tends to a constant at low frequencies. This is because the need to use the measured data to fit for the free parameters in the pulsar timing model hasn't been accounted for. Heuristically the effect of this fitting may be understood by approximating the timing model as a quadratic in time. To motivate the choice of a quadratic notice that a constant offset in the timing residuals is degenerate with mis-estimating the distance to the pulsar, a linear drift in the timing residuals is degenerate with the pulsar spin and a quadratic trend in the timing residuals is degenerate with the pulsar spin-down rate. Expanding the sinusoidal signal at low frequencies ($ft\ll 1$) as a Taylor series in powers of $ft$ allows us to include the effect of fitting for the quadratic timing model (see \citep{MooreTaylorGair2014}) by subtracting off the first three terms in this series (up to and including ${\cal{O}}(ft)^{2}$). The resulting sensitivity curve is shown in Fig.\ \ref{fig:mono_comparison}. This procedure is rather \emph{ad-hoc}, a much more satisfactory way of accounting for the timing model is provided by the Bayesian approach, which we now discuss.

\subsection{Bayesian detection}\label{subsec:bayemono}
The Bayesian procedure for claiming a detection of a signal in the PTA dataset is to consider two competing hypotheses which attempt to describe the measured data, ${\bf{s}}$: the noise hypothesis, $\left\{{\cal{H}}_{n}:{\bf{s}}={\bf{n}}+{\bf{m}}\right\}$; and the signal hypothesis, $\left\{{\cal{H}}_{h}:{\bf{s}}={\bf{n}}+{\bf{m}}+{\bf{h}}\right\}$. Here ${\bf{n}}$ denotes the random noise, ${\bf{m}}$ the timing model and ${\bf{h}}$ the signal; any of ${\bf{n}}$, ${\bf{m}}$ or ${\bf{h}}$ may depend on a number of free parameters, $\vec{\theta}$. For each hypothesis $i\in \left\{n,h\right\}$ the likelihood, ${\cal{L}}_{i}$, may be evaluated for a given choice of the parameters. The evidence for each hypothesis, ${\cal{O}}_{i}$, may be calculated by integrating, or \emph{marginalising}, the likelihood over all of the free parameters in that hypothesis (weighted by any prior knowledge of each parameter). The odds ratio (or Bayes factor) is then defined as the ratio of the evidences ${\cal{B}}={\cal{O}}_{h}/{\cal{O}}_{n}$ and a detection is claimed if the Bayes factor exceeds some predetermined threshold, ${\cal{B}}_{\textrm{th}}$. The threshold Bayes factor is the Bayesian equivalent of the frequentist threshold SNR, and ${\cal{B}}_{\textrm{th}}$ may be chosen to fix a desired false alarm rate.

Here the timing model was taken to be a quadratic in each pulsar, $m_{x}(t)=\alpha_{x}+\beta_{x}t+\gamma_{x}t^{2}$, where $\alpha_{x}$, $\beta_{x}$ and $\gamma_{x}$ are free parameters with flat priors. These parameters are then marginalised over when evaluating the evidence for each hypothesis, this provides a rigorous method for accounting for the loss of sensitivity at low frequencies caused by fitting for the timing model. In reality the timing model is significantly more complex than a simple quadratic; it must account for effects as diverse as the Roemer delay, dispersion of the pulse by the interstellar medium and the orbital motion of the pulsar (if it happens to be in a binary) (\citep{tempo2-1,tempo2-2}). However the simple quadratic timing model considered here captures well the qualitative effect of fitting for the pulsar parameters and has the attractive feature that the free parameters ($\alpha_{x}$, $\beta_{x}$ and $\gamma_{x}$) may be marginalised over analytically.

\begin{figure}[ht]
 \centering
 \includegraphics[trim=0cm 0cm 0cm 0cm, width=0.9\textwidth]{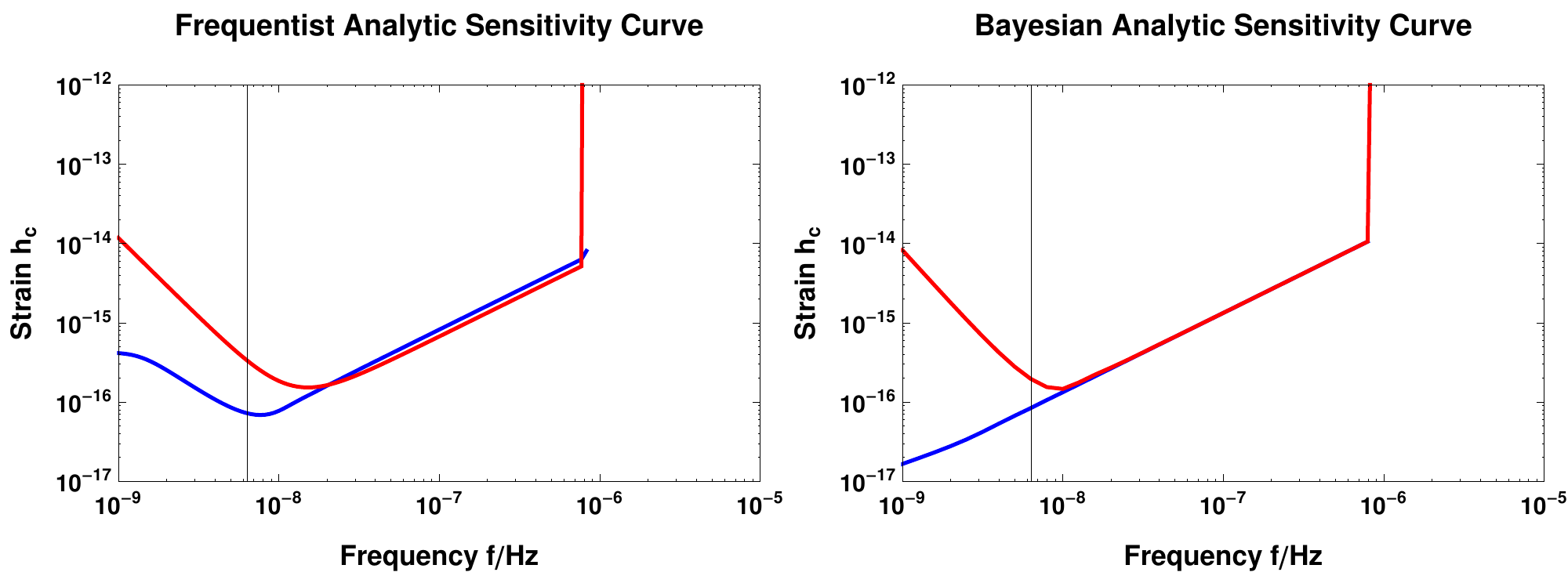}
 \caption{The sensitivity of our canonical PTA to a monochromatic source. The left-hand panel shows the prediction of the frequentist analytic formula in Sec.\ \ref{subsec:freqmono} with (red) and without (black) the effect of fitting for the timing model. The right-hand panel shows the prediction of the Bayesian analytic formula in Sec.\ \ref{subsec:bayemono} with (red) and without (black) the effect of marginalising over the timing model parameters. In both panels it can be seen that the effect of including the timing model is a reduction in sensitivity at low frequencies. (Figure reproduced with modifications from \cite{MooreTaylorGair2014}.)}
 \label{fig:mono_comparison}
\end{figure}

\begin{figure}[ht]
 \centering
 \includegraphics[trim=0cm 0cm 0cm 0cm, width=0.95\textwidth]{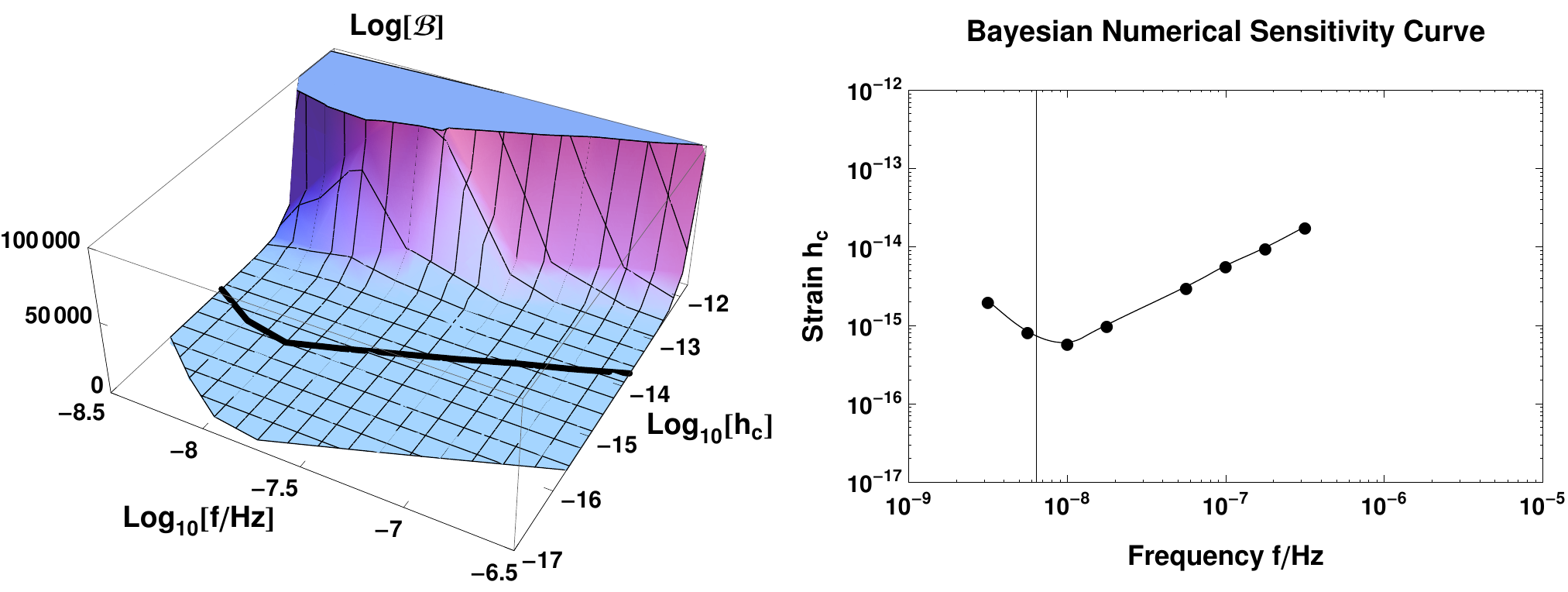}
 \caption{The results of the recovery of numerical injections of a monochromatic signal into the mock dataset. The left-hand panel shows a surface plot of the Bayes factor against the amplitude and frequency of a monochromatic source with the contour ${\cal{B}}={\cal{B}}_{\textrm{th}}$ indicated by a solid black line. The right-hand panel shows the contour ${\cal{B}}={\cal{B}}_{\textrm{th}}$ in a form more comparable to Fig.\ \ref{fig:mono_comparison}. For more detail on the numerical calculations used to produce these plots see \cite{MooreTaylorGair2014}. (Figure reproduced from \cite{MooreTaylorGair2014}.) }
 \label{fig:mono_surface}
\end{figure}

The likelihood functions for both hypotheses may be written in terms of the noise covariance matrix, ${\bf{\Sigma}}_{n}=\sigma\mathds{1}$, where it has been assumed for simplicity that $\sigma_{x}=\sigma,\, \forall x$,
\begin{eqnarray}\label{eq:logLn} &\log {\cal{L}}_{n}(\vec{\Theta})=\log A-\frac{1}{2}\left({\bf{s}}-{\bf{m}}(\vec{\Theta})\right)^{\textrm{T}}{\bf{\Sigma}}_{n}^{-1}\left({\bf{s}}-{\bf{m}}(\vec{\Theta})\right)\;, \\
&\log {\cal{L}}_{h}(\vec{\Theta},\vec{\Psi})=\log A-\frac{1}{2}\left({\bf{s}}-{\bf{m}}(\vec{\Theta})-{\bf{h}}(\vec{\Psi})\right)^{\textrm{T}}{\bf{\Sigma}}_{n}^{-1}\left({\bf{s}}-{\bf{m}}(\vec{\Theta})-{\bf{h}}(\vec{\Psi})\right)\;.\nonumber\end{eqnarray}
The evidence for each hypothesis may be evaluated by marginalising over the free parameters; for the pulsar parameters, $\vec{\Theta}$, flat priors were used, for the source parameters, $\vec{\Psi}$, delta function priors at the true values were used. The choice of delta function priors may seem unreasonable at first, but in reality it is simply a convenient way of imposing a large signal approximation. A detection of GWs is only claimed if the Bayes factor exceeds some threshold, this threshold must be chosen to be large if we are to have high confidence in the detection. In the limit of a large Bayes factor the prior becomes uninformative compared to the data and the value of the Bayes factor obtained in the experiment is independent of the choice of prior. Therefore a delta-function prior is chosen to ease the analytic calculations. For a further discussion of this point see \cite{MooreTaylorGair2014}. The evidence for hypothesis ${\cal{H}}_{i}$ is given by
\begin{equation} {\cal{O}}_{i}({\bf{s}})=\int\textrm{d}\vec{\lambda}_{i}\;{\cal{L}}_{i}({\bf{s}},\vec{\lambda}_{i})P_{i}(\vec{\lambda}_{i}) \; ,\quad \textrm{where}\;\lambda_{n}=\vec{\theta}\; \textrm{and}\; \lambda_{h}=\left\{\vec{\theta},\vec{\psi}\right\}\,.\end{equation}
The Bayes factor, ${\cal{B}}$, may now be evaluated and the expectation over noise realisations calculated,
\begin{equation}\label{eq:final} \overline{{\cal{B}}}=\int\textrm{d}{\bf{n}}\;P({\bf{n}}){\cal{B}}\quad \textrm{where}\; P({\bf{n}})=\frac{\exp\left(-\frac{1}{2}{\bf{n}}^{\textrm{T}}{\bf{\Sigma}}_{n}^{-1}{\bf{n}}\right)}{\sqrt{(2\pi)^{N_{p}T/\delta t}\textrm{det}\left({\bf{\Sigma}}_{n}\right)}}\; .\end{equation}
The integral in Eq.\ \ref{eq:final} can be performed analytically and inverted to give an analytic expression for the  minimum detectable amplitude as a function of frequency. This expression is too lengthy to reproduce here (see \cite{MooreTaylorGair2014}) but is plotted in Fig.\ \ref{fig:mono_comparison}. Both of the analytic curves shown in Fig.\ \ref{fig:mono_comparison} should be compared with the results of the numerical injections and recovery shown in Fig. \ref{fig:mono_surface}.

\section{Sensitivity to a stochastic background of GWs}\label{sec:stoch}
A stochastic background of gravitational waves is completely specified by the power spectral density, $S_{h}(f)$, which is usually taken to be a power law in frequency,
\begin{equation}\label{eq:OverlapReductionFunction} \left<\tilde{h}_{x}(f)\tilde{h}^{*}_{y}(f')\right>=\frac{1}{2}\delta (f-f') \Gamma_{xy} S_{h}(f)\;, \quad \textrm{where}\; S_{h}(f)=\frac{A^{2}}{12\pi^{2}f_{0}^{3}}\left(\frac{f}{f_{0}}\right)^{-\gamma}\; .\end{equation}
The constant $\Gamma_{xy}$ is called the overlap reduction function and, for an isotropic background, depends only on the relative sky positions of pulsars $x$ and $y$. As can be seen from Eq.\ \ref{eq:OverlapReductionFunction} the signal has contributions from all frequencies, in contrast to the monochromatic signal in Sec.\ \ref{sec:mono}. When calculating the sensitivity curve it no longer makes sense to calculate the minimum detectable amplitude as a function of frequency, $h_{c}(f)$. Instead the minimum detectable amplitude of the background in terms of the power-law slope may be determined, $A(\gamma)$. Unfortunately the interpretation of this quantity is somewhat opaque, so an approach suggested by \cite{thraneromano2013} can be used. The parameter $\gamma$ is gridded with $n$ values selected between some large-negative and large-positive limits. For each value of $\gamma$ the value $A(\gamma)$ may be calculated, giving the set of ordered pairs $(\gamma_{i},A_{i})$ for $i\in \left\{1,2,\ldots n\right\}$. For each pair the power-law curve for characteristic strain, $h_{c}=A(\gamma)\left(f/f_{0}\right)^{(\gamma-3)/2}$, may be drawn. This gives a set of power-law curves (straight lines on a log-log plot) for characteristic strain, the envelope of this set of curves is then interpreted as the \emph{power-law-integrated} sensitivity curve \citep{thraneromano2013}, See Figs.\ \ref{fig:stoch_comparison} and \ref{fig:stoch_surface}.

\subsection{Frequentist detection}
Similar calculations to those performed in Sec.\ \ref{subsec:freqmono} may now be performed for the stochastic background. The principal difference is that in the expression for the filter function $\tilde{{\bf{K}}}(f)$ (Eq.\ \ref{eq:statistic}) the product $\tilde{h}_{x}(f)\tilde{h}_{y}(f)$ is replaced by the expectation value $\left<\right.$$\tilde{h}_{x}(f)\tilde{h}_{y}(f)$$\left.\right>$. The resulting expression for the SNR is (see \citep{MooreTaylorGair2014})
\begin{equation}\label{eq:snrback}  \rho^{2}= \sum_{x>y}\sum_{y}8T\int\textrm{d}f\;\frac{\Gamma_{xy}^{2}S_{h}^{2}(f)}{S^{2}_{n}(f)} \;.\end{equation}
Setting $\varrho=\varrho_{\textrm{th}}$ in Eq.\ \ref{eq:snrback} and inverting gives the expression $A(\gamma)$ \citep{MooreTaylorGair2014} and the power-law-integrated sensitivity curve may be drawn following the procedure outlined above, see Fig.\ \ref{fig:stoch_comparison}.

\subsection{Bayesian detection}
\begin{figure}[ht]
 \centering
 \includegraphics[trim=0cm 0cm 0cm 0cm, width=0.9\textwidth]{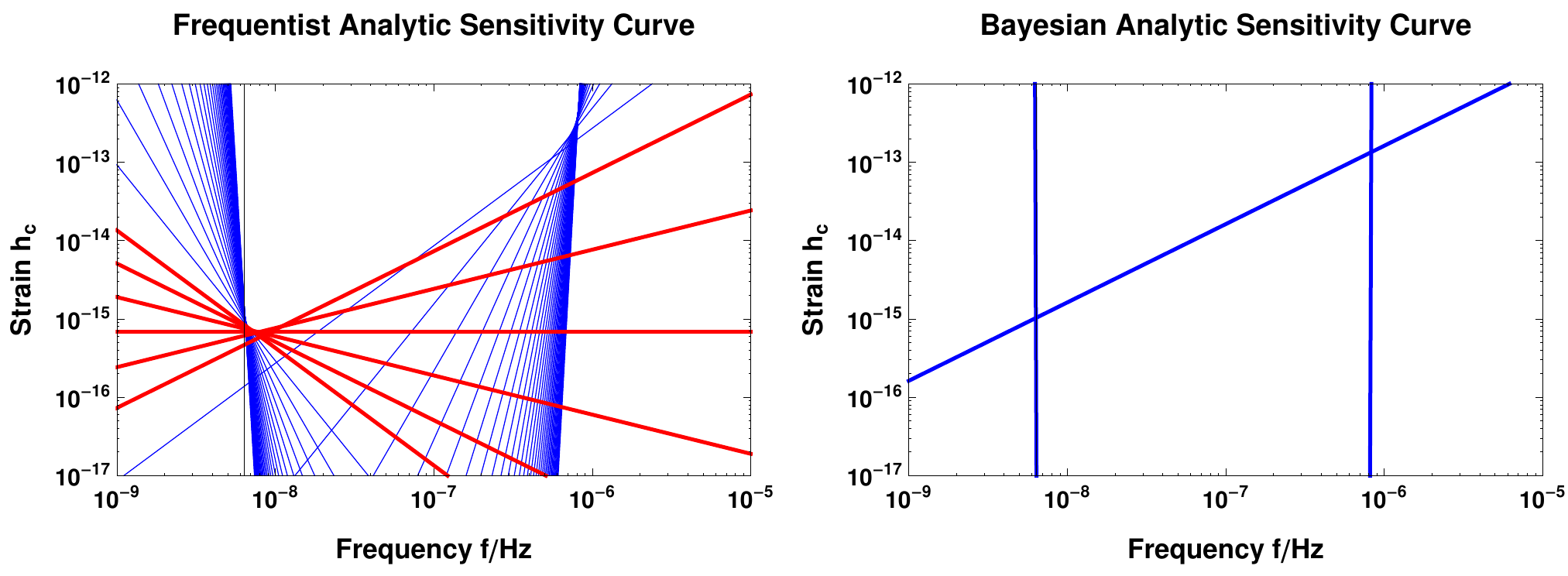}
 \caption{The sensitivity of our canonical PTA to a stochastic background. The left-hand panel shows the prediction of the frequentist analytic formula in Sec.\ \ref{subsec:freqmono}, the curves drawn in red are for comparison with Fig.\ \ref{fig:stoch_surface}. The right-hand panel shows the prediction of the Bayesian analytic formula in Sec.\ \ref{subsec:bayemono}. (Figure reproduced from \cite{MooreTaylorGair2014}.)}
 \label{fig:stoch_comparison}
\end{figure}

\begin{figure}[ht]
 \centering
 \includegraphics[trim=0cm 0cm 0cm 0cm, width=0.95\textwidth]{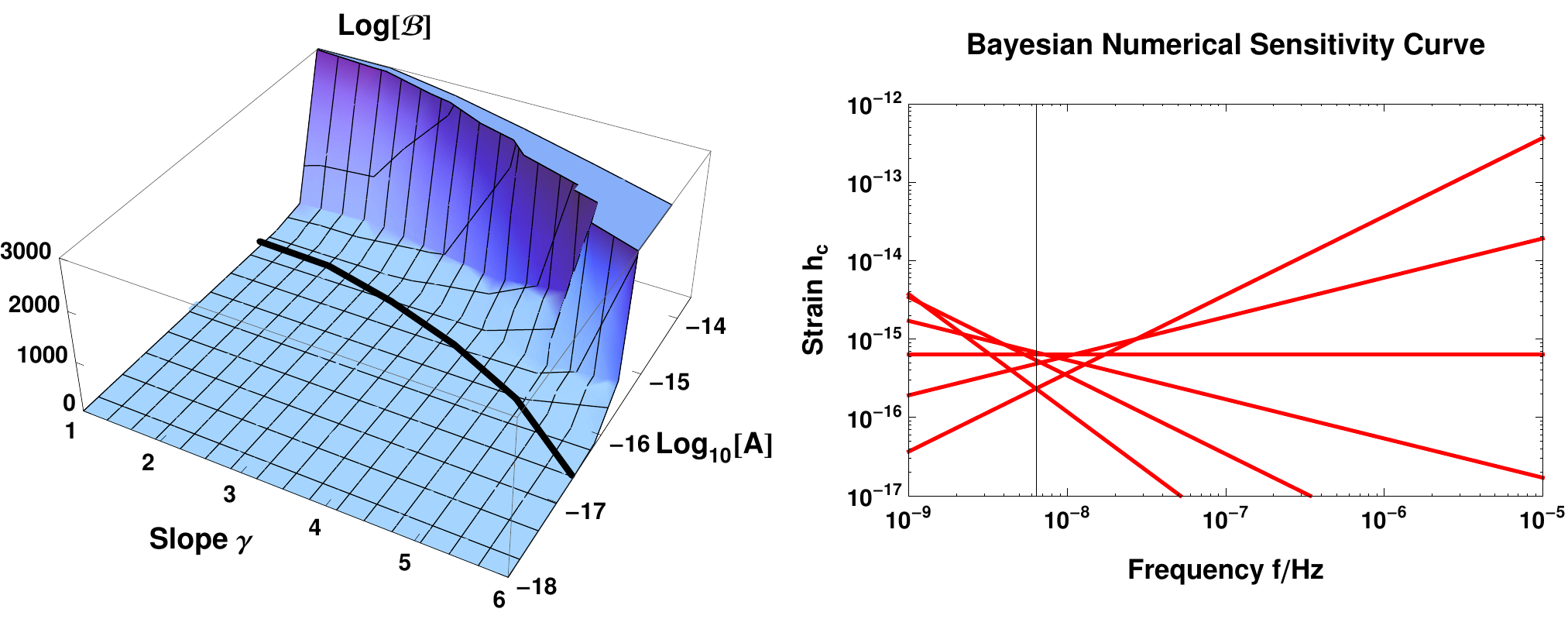}
 \caption{The results of the recovery of numerical injections of a stochastic background signal into the mock dataset. The left-hand panel shows a surface plot of the Bayes factor against the amplitude and power-law index of a stochastic background with the contour ${\cal{B}}={\cal{B}}_{\textrm{th}}$ indicated by a solid black line. The right-hand panel shows a power-law-integrated sensitivity curve formed by using points off the contour ${\cal{B}}={\cal{B}}_{\textrm{th}}$. The range of values of the power-law index was limited for reasons of numerical stability however this plot may be compared with the curves drawn in red in Fig.\ \ref{fig:stoch_comparison}. For more detail on the numerical calculations used to produce these plots see \cite{MooreTaylorGair2014}. (Figure reproduced from \cite{MooreTaylorGair2014}.)}
 \label{fig:stoch_surface}
\end{figure}

Similar calculations to those performed in Sec.\ \ref{subsec:bayemono} may now be performed for the stochastic background. This yields an analytic expression for the expectation value of the Bayes factor, $\overline{{\cal{B}}}$. Setting $\overline{{\cal{B}}}={\cal{B}}_{\textrm{th}}$ and inverting gives an analytic expression for $A(\gamma)$. The analytic expressions are too lengthy to reproduce here, but can be found in \cite{MooreTaylorGair2014}. Using the expression $A(\gamma)$ the power-law-integrated sensitivity curve may be drawn following the procedure outlined above, see Fig.\ \ref{fig:stoch_comparison}.

\section{Conclusions}\label{sec:conclusions}
The sensitivity of a PTA depends not only on the properties of the PTA itself, but also on the properties of the source being observed. This statement is also true for more traditional interferometric detectors such as aLIGO and eLISA (\citep{2010CQGra..27h4006H,2012CQGra..29l4016A}) however the effect is particularly important for PTAs because their sensitivity is determined by the sampling properties of the dataset rather than properties of the noise. For example, in the case of LIGO the lowest frequency which may be detected is determined to a large extent by the properties of the seismic noise, whereas in the case of PTAs it is determined by the total length of the observations. 

The sensitivity curve of a canonical PTA was evaluated for both a monochromatic source and a stochastic background, and this was done in both the frequentist and Bayesian frameworks, with consistent results found. The analytic calculations were also verified by numerical injections into a mock dataset. In the simplified case of regularly sampled data with white, Gaussian, uncorrelated noise, simple analytic expressions for the sensitivity of a PTA have been derived. It is hoped that these expressions will be useful going forward in order to assess the prospects for detection and astronomical inference using PTA observations.

\subsection*{Acknowledgments}
The author is supported by the STFC. This talk was a partial summary of work done in collaboration with S. R. Taylor and J. R. Gair. The author would like to thank the organisers of the 10th international LISA symposium for the opportunity to give this talk.

\bibliographystyle{iopart-num}
\bibliography{bibliography}

\providecommand{\newblock}{}
\begin{thebibliography}{10}
\expandafter\ifx\csname url\endcsname\relax
  \def\url#1{{\tt #1}}\fi
\expandafter\ifx\csname urlprefix\endcsname\relax\def\urlprefix{URL }\fi
\providecommand{\eprint}[2][]{\url{#2}}

\bibitem{jaffe-backer-2003}
{Jaffe} A~H and {Backer} D~C 2003 {\em \apj\/} {\bf 583} 616--631
  (\textit{Preprint} \eprint{arXiv:astro-ph/0210148})

\bibitem{wyithe-loeb-2003}
{Wyithe} J~S~B and {Loeb} A 2003 {\em \apj\/} {\bf 590} 691--706
  (\textit{Preprint} \eprint{arXiv:astro-ph/0211556})

\bibitem{sesana-vecchio-colacino-2008}
{Sesana} A, {Vecchio} A and {Colacino} C~N 2008 {\em \mnras\/} {\bf 390}
  192--209 (\textit{Preprint} \eprint{0804.4476})

\bibitem{sesana-vecchio-volonteri-2009}
{Sesana} A, {Vecchio} A and {Volonteri} M 2009 {\em \mnras\/} {\bf 394}
  2255--2265 (\textit{Preprint} \eprint{0809.3412})

\bibitem{sesana-vecchio-2010}
{Sesana} A and {Vecchio} A 2010 {\em Classical and Quantum Gravity\/} {\bf 27}
  084016 (\textit{Preprint} \eprint{1001.3161})

\bibitem{olmez-2010}
{{\"O}lmez} S, {Mandic} V and {Siemens} X 2010 {\em \prd\/} {\bf 81} 104028
  (\textit{Preprint} \eprint{1004.0890})

\bibitem{grishchuk-1976}
{Grishchuk} L~P 1976 {\em Pis ma Zhurnal Eksperimental noi i Teoreticheskoi
  Fiziki\/} {\bf 23} 326--330

\bibitem{grishchuk-2005}
{Grishchuk} L~P 2005 {\em Physics Uspekhi\/} {\bf 48} 1235--1247
  (\textit{Preprint} \eprint{arXiv:gr-qc/0504018})

\bibitem{eptareview2013}
{Kramer} M and {Champion} D~J 2013 {\em Classical and Quantum Gravity\/} {\bf
  30} 224009

\bibitem{parkesreview2013}
{Hobbs} G 2013 {\em Classical and Quantum Gravity\/} {\bf 30} 224007
  (\textit{Preprint} \eprint{1307.2629})

\bibitem{nanogravreview2013}
{McLaughlin} M~A 2013 {\em Classical and Quantum Gravity\/} {\bf 30} 224008
  (\textit{Preprint} \eprint{1310.0758})

\bibitem{iptareview2013}
{Manchester} R~N and {IPTA} 2013 {\em Classical and Quantum Gravity\/} {\bf 30}
  224010

\bibitem{MooreTaylorGair2014}
{Moore} C~J, {Taylor} S~R and {Gair} J~R 2014 {\em ArXiv e-prints\/}
  (\textit{Preprint} \eprint{1406.5199})

\bibitem{tempo2-1}
{Hobbs} G~B, {Edwards} R~T and {Manchester} R~N 2006 {\em \mnras\/} {\bf 369}
  655--672 (\textit{Preprint} \eprint{arXiv:astro-ph/0603381})

\bibitem{tempo2-2}
{Edwards} R~T, {Hobbs} G~B and {Manchester} R~N 2006 {\em \mnras\/} {\bf 372}
  1549--1574 (\textit{Preprint} \eprint{arXiv:astro-ph/0607664})

\bibitem{thraneromano2013}
{Thrane} E and {Romano} J~D 2013 {\em \prd\/} {\bf 88} 124032
  (\textit{Preprint} \eprint{1310.5300})

\bibitem{2010CQGra..27h4006H}
{Harry} G~M and {LIGO Scientific Collaboration} 2010 {\em Classical and Quantum
  Gravity\/} {\bf 27} 084006

\bibitem{2012CQGra..29l4016A}
{Amaro-Seoane} P, {Aoudia} S, {Babak} S, {Bin{\'e}truy} P, {Berti} E,
  {Boh{\'e}} A, {Caprini} C, {Colpi} M, {Cornish} N~J, {Danzmann} K, {Dufaux}
  J~F, {Gair} J, {Jennrich} O, {Jetzer} P, {Klein} A, {Lang} R~N, {Lobo} A,
  {Littenberg} T, {McWilliams} S~T, {Nelemans} G, {Petiteau} A, {Porter} E~K,
  {Schutz} B~F, {Sesana} A, {Stebbins} R, {Sumner} T, {Vallisneri} M, {Vitale}
  S, {Volonteri} M and {Ward} H 2012 {\em cqg\/} {\bf 29} 124016
  (\textit{Preprint} \eprint{1202.0839})

\end{thebibliography}
\end{document}